\documentclass[10pt,conference]{IEEEtran}
\IEEEoverridecommandlockouts

\usepackage{cite}
\usepackage{mdframed}
\usepackage{bm}
\usepackage{amsmath,amssymb,amsfonts}
\usepackage{subfigure}
\usepackage{url}
\usepackage{multirow}
\usepackage{graphicx}
\usepackage{tcolorbox}
\usepackage{bbm}
\usepackage{rotating}

\usepackage{xcolor}
\definecolor{LightGray}{gray}{0.9}

\usepackage{hyperref}
\hypersetup{backref=true, pagebackref=true, hyperindex=true, colorlinks=true, breaklinks=true, urlcolor= black, linkcolor= blue, bookmarks=true, bookmarksopen=false, filecolor=black, citecolor=black, linkbordercolor=blue}
\usepackage{colortbl}
\usepackage{cleveref}

\def\BibTeX{{\rm B\kern-.05em{\sc i\kern-.025em b}\kern-.08em
    T\kern-.1667em\lower.7ex\hbox{E}\kern-.125emX}}

\begin{document}
\title{Domain Adaptive Code Completion via Language Models and Decoupled Domain Databases}

\author{
\IEEEauthorblockN{Ze Tang\textsuperscript{1}, Jidong Ge\textsuperscript{1}\IEEEauthorrefmark{1}, Shangqing Liu\textsuperscript{2}\IEEEauthorrefmark{1}, Tingwei Zhu\textsuperscript{1}, Tongtong Xu\textsuperscript{3}, Liguo Huang\textsuperscript{4}, Bin Luo\textsuperscript{1}} 

\IEEEauthorblockA{
		\textsuperscript{1}National Key Laboratory for Novel Software Technology, Nanjing University, China\\
		\textsuperscript{2}Nanyang Technological University, Singapore\\
        \textsuperscript{3}Huawei Software Engineering Application Technology, China\\
        \textsuperscript{4}Department of Computer Science, Southern Methodist University, USA\\
}

\IEEEauthorblockA{
		\{zetang, tingweizhu33\}@smail.nju.edu.cn, \{gjd, luobin\}@nju.edu.cn, liu.shangqing@ntu.edu.sg,\\ xutongtong9@huawei.com, lghuang@lyle.smu.edu
}

\thanks{\IEEEauthorrefmark{1}Corresponding author.}

}

\maketitle

\begin{abstract}
Large Language Models (LLMs) have demonstrated remarkable performance in code completion. However, due to the lack of domain-specific knowledge, they may not be optimal in completing code that requires intensive domain knowledge for example completing the library names. 
Although there are several works that have confirmed the effectiveness of fine-tuning techniques to adapt language models for code completion in specific domains. They are limited by the need for constant fine-tuning of the model when the project is in constant iteration.

To address this limitation, in this paper, we propose $k$NM-LM, a retrieval-augmented language model (R-LM), that integrates domain knowledge into language models without fine-tuning. Different from previous techniques, our approach is able to automatically adapt to different language models and domains. Specifically, it utilizes the in-domain code to build the retrieval-based database decoupled from LM, and then combines it with LM through Bayesian inference to complete the code. The extensive experiments on the completion of intra-project and intra-scenario have confirmed that $k$NM-LM  brings about appreciable enhancements when compared to CodeGPT and UnixCoder. A deep analysis of our tool including the responding speed, storage usage, specific type code completion, and API invocation completion has confirmed that $k$NM-LM provides satisfactory performance, which renders it highly appropriate for domain adaptive code completion. Furthermore, our approach operates without the requirement for direct access to the language model's parameters. As a result, it can seamlessly integrate with black-box code completion models, making it easy to integrate our approach as a plugin to further enhance the performance of these models.

\end{abstract}

\begin{IEEEkeywords}
domain adaptive code completion, retrieval-augment language model
\end{IEEEkeywords}

\section{Introduction}
\label{sec:introduction}
Large language models (LLM)~\cite{codexglue,uniXcoder,plbart,niu2022spt,t5}, have achieved state-of-art performance in code completion, and some of them have been successfully used as the auto-completion plugin (e.g. GitHub Copilot~\cite{copilot} and ChatGPT~\cite{chatgpt}) in modern Integrated Development Environment (IDE). Nevertheless, code in IDE has distinctive domain-specific features, such as imported third-party libraries and intra-project references. These features can vary significantly across domains and are continuously updated. Thus, LLMs, which are developed for completing general code, may not perform well when applied to new domains such as personal projects, as noted in prior works~\cite{structuralLocality,project_miner,zancert}.

A standard practice to adapt a pre-trained model to a new domain (i.e., domain adaptation) is model fine-tuning~\cite{keepPretraining2020}. By fine-tuning a language model on a new domain, the model's performance on that domain can be improved without having to retrain the entire model from scratch. Nevertheless, fine-tuning can be infeasible for large language models. For instance, language models such as GPT-3 or GPT-4, are often deployed as black-box systems. The black-box limitation renders the \textbf{parameters inaccessible} to users, thereby preventing them from fine-tuning. More importantly, domain-specific features in code are subject to \textbf{frequent changes}. This is due to the continuous development of projects and the incessant updates of third-party libraries utilized in the code. Consequently, the employment of fine-tuning for domain-intensive code completion may not be a feasible option.

In recent years, a series of retrieval-augmented language models (R-LM)~\cite{guu2020retrieval,ReACC,parvez2021retrieval,knnlm} have emerged to address this challenge. The key ingredient of R-LMs is their ability to utilize the domain database at test time without having to rely on the information encoded in the model's weights only. In these models, the retrieval component first searches for nearest neighbor examples in an external datastore (e.g., code from the same project); then, the base model references these examples during the prediction. One prominent example of such a retrieval-based model is $k$-nearest neighbors language model ($k$NN-LM), which predicts a token by linearly interpolating the base LM's output with a non-parametric nearest neighbor distribution. This distribution is constructed by searching for the k-nearest neighbors ($k$NN) in the datastore and weighing them according to their distance to the current test context. Notably, $k$NN-LM requires a large datastore that stores each token in the domain and is sensitive to the manual-selected interpolated weight for soft voting, making them challenging to apply to code completion. Developers may not have the resources or patience to find suitable hyper-parameters for their own code.

In this paper, we introduce a plug-and-play auto-completion solution that does not need careful configuration, the $k$-nearest mistakes language model ($k$NM-LM). Unlike prior R-LMs, $k$NM-LM retrieves information from a separate database that is decoupled from the language model. This is because the weights in the language model already contain some content knowledge about the program language, such as grammar and built-in methods. Thus, storing this information redundantly in the database would be both wasteful and meaningless. To overcome this challenge, we split the domain knowledge into two subsets: what the language model knows and what it doesn't. The latter is saved as a decoupled database, which is a subset of the domain database that stores only tokens that cannot be correctly predicted by the language model (i.e., mistakes collection). Decoupling offers two primary benefits. Firstly, the decoupled database requires less storage and search resources. Secondly, the decoupled database and language model are two distinct and non-intersecting systems. As a result, we do not need to manually select the interpolated weight but can leverage statistical approaches, such as Bayesian inference, to automatically combine the database and language model from a statistical perspective.

Specifically, we start by creating a datastore at the token level. This datastore only contains code tokens that the language model fails to predict correctly. We then retrieve tokens from the datastore that are most like the code being completed and normalize them into a distribution. Next, we adopt the error rate of the language model on the original domain database as the prior probability and assess whether the language model can predict the code tokens prior to the completion position as new evidence (likelihood probability). The posterior probability is then calculated and utilized to merge the distributions obtained from the datastore and the language model.

We conducted an evaluation of $k$NM-LM on intra-project and intra-scenario code completion, revealing the performance improvement in both CodeGPT and UnixCoder through our proposed approach. We further performed validation on three key aspects that could affect the user experience when using $k$NM-LM as a code completion plugin. These included completion speed and space utilization, the performance in completing specific code types and completing specific API calls from third-party libraries. Our experimental results demonstrate that $k$NM-LM outperforms several baseline models and is better suited for domain adaptive code completion than other R-LMs. Additionally, $k$NM-LM does not require access to the language model's weights, making it suitable for situations where only black-box access to the language model is available. In summary, the primary contributions of this paper are as follows:
\begin{itemize}
\item We propose $k$NM-LM, a retrieval-augmented language model (R-LM), that can be used for domain adaptive code completion without fine-tuning. All code, data, and results can be found at our anonymous repository\footnote{\url{https://github.com/zetang94/ASE2023_kNM-LM}}.
\item Different from previous R-LMs, $k$NM-LM retrieves from a decoupled database and uses Bayesian inference to interpolate between the database and language model.
\item Experimental results on intra-project and intra-scenario code completion demonstrate that $k$NM-LM improves the performance of both CodeGPT and UniXcoder.
\item We also investigate potential issues that may impact user experience when using $k$NM-LM as a code completion plugin. Our results indicate that $k$NM-LM performs satisfactorily across all considered factors.
\end{itemize}

\begin{figure*}[!t]
    \centering
    \includegraphics[width=\linewidth]{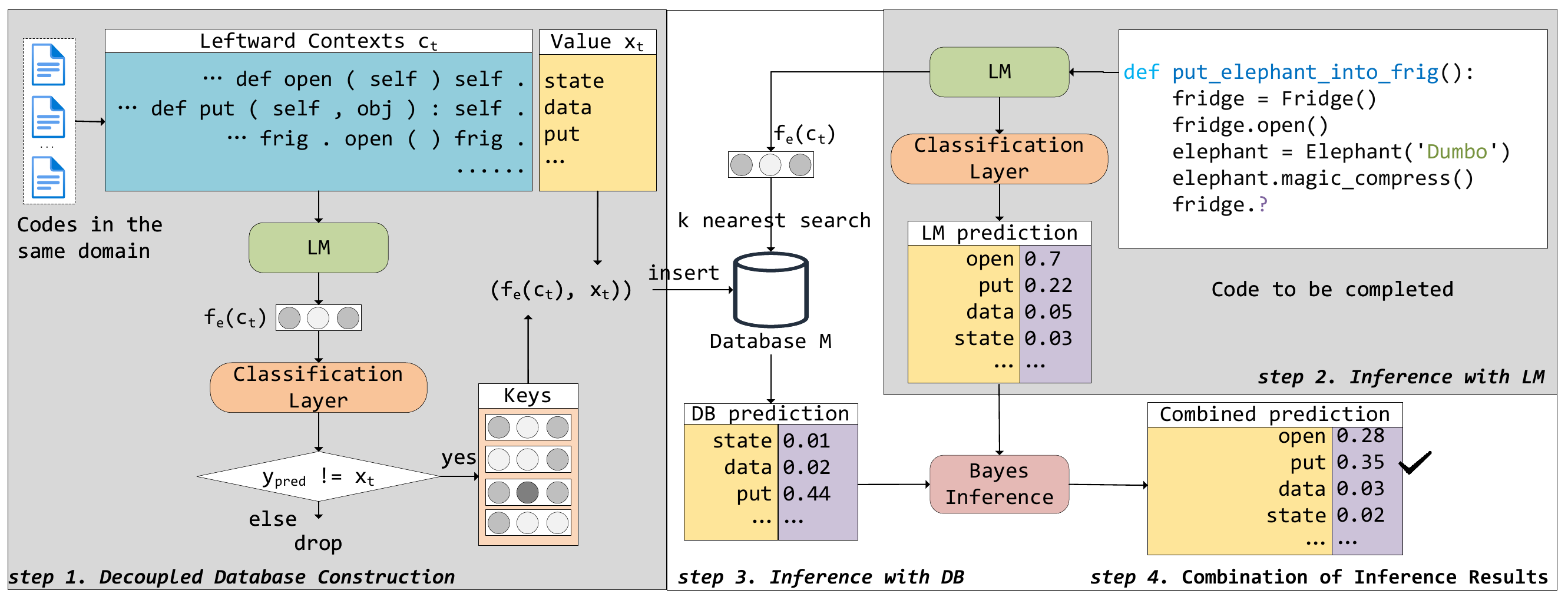}
    \caption{The primary workflow of our proposed $k$NM-LM. A grey box indicates that this part needs to access the (black-box) language model.}
    \label{fig:model_structure}
\end{figure*}
\section{Background}
\subsection{$k$-Nearest Neighbors Language Model}
$k$-Nearest Neighbors Language Model ($k$NN-LM)~\cite{knnlm} is a retrieval-augmented language model (R-LM), in which
uses a nearest neighbor retrieval mechanism to augment the pre-trained language model, without any additional fine-tuning. Given a sequence of tokens $c_t = (x_1, x_2, \dots, x_{t-1})$, autoregressive language models, such as GPT-3, estimate $p_{LM}(y|c_t)$, the probability distribution over the next token. 

\noindent \textbf{Datastore.} Let $f_e(\cdot)$ be the encoding function that maps a context $c_t$ to a fixed-length vector representation computed by the pre-trained language model. First, $k$NN-LM generates the key-value datastore by running the encoding function $f_e(\cdot)$ over a corpus $\mathcal{D}$. Typically, each key $k_t$ is the vector representation $f_e(c_t)$ of the context $c_t$, and each value $v_t$ is the next token $x_t$ for the context $c_t$, as:
\begin{equation}
(K, V) = \{ (f_e(c_t), x_t) | (c_t, x_t) \in \mathcal{D}\}
\end{equation}

\noindent \textbf{Inference.} At test time, given the input context $c_t$, the pre-trained language model generates the probability distribution over the next token $p_{LM}(y|c_t)$ and the context representation $f_e(c_t)$. $k$NN-LM queries the datastore to retrieve $k$-nearest keys according to a vector distance function $d(\cdot, \cdot)$. Then, the model computes a softmax over the (negative) distances, which gives a distribution over the next token:
\begin{equation}
   p_{kNN}(y|c_t) \propto \sum_{(k_t,v_t) \in \mathcal{D}} \mathbbm{1}_{y= v_t}\exp(-d(k_t, f_e(c_t))) 
\end{equation}
The prediction is then interpolated with the prediction from the language model:
\begin{equation}\label{equation:soft_voting}
    p(y|c_t)=\lambda p_{kNN}(y|c_t) + (1-\lambda) p_{LM}(y|c_t)
\end{equation}
where $\lambda$ is a hyper-parameter that ranges from $0$ to $1$ and needs to be carefully selected.

\subsection{Bayesian Inference}
Bayesian inference is a fundamental concept in probability theory that allows us to update our beliefs about an event considering new evidence or information. It states that the probability of a hypothesis or event (A) given some observed evidence (C) can be calculated using Bayes' rule:
\begin{equation}
    p(A|C) \propto p(A) \times p(C|A)
\end{equation}
where $p(A)$ is the prior probability of $A$, $p(C|A)$ is the probability of observing $C$ given that $A$ is true (the likelihood), and $p(A|C)$ is the posterior probability of $A$ given $C$. Bayesians consider $\theta=p(A)$ and use $p(\theta)$ to describe the distribution of $p(A)$. Specifically, if $p(\theta) = Beta(a, b)$, and event $C$ indicates that $A$ occurs $\alpha$ times in $N$ new observations. Then we have the following conclusions:
\begin{equation}
\begin{aligned}
&\text{\textbf{Likelihood}:} &&p(C|\theta) = \theta^{(\alpha-1)}(1-\theta)^{(N-\alpha-1)}\\
&\text{\textbf{Posterior}:} &&p(\theta|C) = Beta(a+\alpha, b+N-\alpha)\\
&\text{\textbf{Expectation}:} &&p(A|C) = E(p(\theta|C)) = \frac{(a+\alpha)}{(a+b+N)}
\end{aligned}
\end{equation}
\section{Approach}
\label{sec:approach}
The code completion task aims to predict the next token or the next line of code given a known code sequence $c_t=(x_1,x_2, \dots, x_{t-1})$. For domain-adaptive code completion, we assume that some in-domain code can be obtained (e.g., code already written in current project or code from other projects that are developed for the same application scenario). Retrieval-augmented language models (R-LMs) can be employed to enhance the performance of general (black-box) LMs for completing domain-intensive code. The primary workflow of our proposed method, $k$-nearest mistakes language model ($k$NM-LM), is illustrated in Fig.~\ref{fig:model_structure}.

Initially, we analyze the R-LMs from a statistical perspective (\cref{sec:statistical_perspective}), discovering that manual selection of combination coefficient in previous R-LMs can be avoided by decoupling the retrieval module and LM. Next, we discuss the construction of the decoupled retrieval database (\cref{sec: decoupled_database}). During the inference phase, we separately predict the next token probability using the LM (\cref{sec: inference_lm}) and the retrieval module (\cref{sec: inference_db}). Finally, we employ Bayesian inference to combine the inferences (\cref{sec: combined_inference}).

\subsection{Analysis of R-LM from a Statistical Perspective}
\label{sec:statistical_perspective}
We initiate our discussion by defining two distinct events associated with predicting the next token using LM:

\noindent \textbf{Event} $E$: The LM correctly predicts the next token of $c_t$.
\begin{itemize}[leftmargin=*]
\item Set notation: $E=\{x_t|\arg\max(p_{LM}(y|c_t)) = x_t\}$
\item Description: The predicted token $y$ with the highest probability  matches the ground truth $x_t$.
\end{itemize}

\noindent \textbf{Event} $E'$: The LM incorrectly predicts the next token of $c_t$.
\begin{itemize}[leftmargin=*]
\item Set notation: $E'=\{x_t|\arg\max(p_{LM}(y|c_t)) \neq x_t\}$
\item Description: The predicted token $y$ with the highest probability  does not match the ground truth $x_t$.
\end{itemize}
The primary goal of retrieval-augmented approaches is to enhance the base LM's performance through error correction. Ideally, for tokens belonging to Event $E$, the retrieval module should not impact the final prediction, allowing the predictions from the LM to be utilized directly. Conversely, for tokens belonging to Event $E'$, the retrieval module aims to ``correct'' the erroneous prediction by searching for previously similar mistakes made by the LM and using the corresponding ground truths as predictions. The idea is similar to the way that humans use a ``collection of mistakes'' to remind themselves not to make the same mistake twice. We can use the total probability theorem to formulate the retrieval-argument process, as:
\begin{equation}\label{eq:total_probability}
\begin{aligned}
p(y|c_t) &=p(y|c_t, E' \cup E)\\
&=p(y|c_t, E')p(E'|c_t) + p(y|c_t, E)p(E|c_t) \\
&=p_{Retrieval}(y|c_t)p(E'|c_t)+p_{LM}(y|c_t)p(E|c_t)
\end{aligned}
\end{equation}
From Equation~\ref{eq:total_probability}, it is essential to decouple the retrieval module's abilities from the LM, as they predict tokens that belong to distinct events. Previous R-LMs do not decouple the retrieval module from the LM, necessitating the use of an ensemble strategy~\cite{ensembleApproach} to combine their strengths. However, the ensemble strategy requires manual and empirical determination of the combination coefficient. Inappropriate coefficient may even lead to a decline in performance. By decoupling the retrieval module, the combination coefficient $p(E'|c_t)$ can be computed mathematically. Moreover, $p(E'|c_t)$ is associated with both the domain-specific context and the LM, facilitating the adaptation of code completion tasks with different LMs across various domains. In the subsequent section, we will discuss the construction of the decoupled database.

\subsection{Step 1. Decoupled Database Construction}
\label{sec: decoupled_database}
Given the known in-domain code $\mathcal{D}$, we build the decoupled database $\mathcal{M}$ as a key-value store. The values consist of tokens associated with event $E'$, while the corresponding keys are the embedded vectors of their preceding contexts:
\begin{equation}
\mathcal{M} = \{ (f_e(c_t), x_t)| (c_t, x_t) \in \mathcal{D}, x_t \in E'\}
\end{equation}
Here, $f_e(\cdot)$ denotes a function that maps a code sequence to a fixed-length representation with dimension $d$. In CodeGPT, for example, $f_e(\cdot)$ might be the output of the last self-attention layer. It is worth noting that the decoupled database is considerably smaller than the one constructed in $k$NN-LM, which saves storage space and search time. The reduction ratio is inversely proportional to the LM's error rate $err$ in $\mathcal{D}$, as:
\begin{equation}
\label{eq:prior_prob}
p(E') = err = \frac{||\mathcal{M}||}{||\mathcal{D}||} 
\end{equation}
$||\cdot||$ signifies the set size. Importantly, the error rate also represents the probability of event $E'$ occurring in $\mathcal{D}$. Assuming that the code to be completed are independent and identically distributed (i.i.d.) with code in $\mathcal{D}$, we can use the error rate as the prior probability $p(E')$ at inference time.

\subsection{Step 2. Inference with LM}
\label{sec: inference_lm}
Language models are treated as a black-box system in our approach, as we do not need to access their parameters. We input the code context $c_t$ into the LM and obtain the probability of the next token $p_{LM}(y|c_t)$. Additionally, we need to acquire the embedded context vector $f_e(c_t)$. All of this can be achieved by utilizing the corresponding APIs provided by the LM. For example, GPT-3 offers APIs that support embedding context~\cite{gpt_api_emb} and obtaining the probabilities of next tokens based on the context~\cite{gpt_api_completion}. After obtaining the embedded code context, we use it as the key to search the database.

\subsection{Step 3. Inference with DB}
\label{sec: inference_db}
We follow Khandelwal et al.~\cite{knnlm} to generate the probability of the next token from the datastore. With the embedded context vector $f_e(c_t)$ generated by the LM, the retrieval module searches for keys similar to $f_e(c_t)$ in the datastore using $k$-nearest neighbors search. Similarity is defined by the vector distance $d(\cdot, \cdot)$ (we employ Euclidean distance in our experiments). The model then computes the distribution over the retrieved results as follows:
\begin{equation}
p_{kNM}(y|c_t) \propto \sum_{(k_i,v_i) \in \mathcal{M}} \mathbbm{1}_{y= v_i}\exp(-d(k_i, f_e(c_t))) \
\end{equation}
$\mathbbm{1}$ is an indicator function, which aggregates probability mass for each vocabulary item across all its occurrences in the retrieved targets, and probabilities for items not appearing in the retrieved targets are set to zero.

\begin{figure}[!t]
    \centering
    \includegraphics[width=\linewidth]{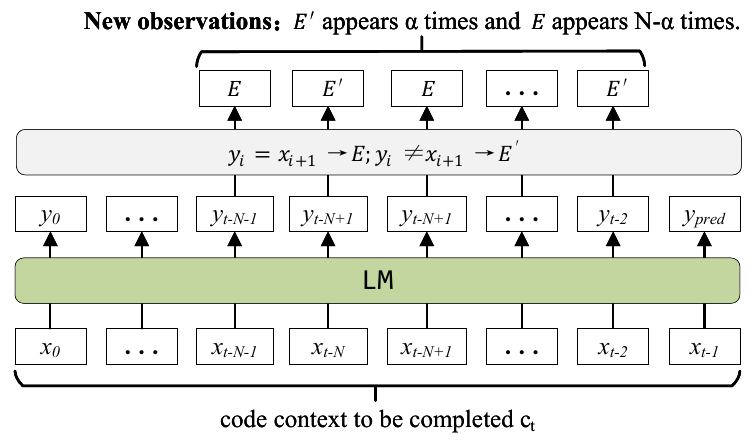}
      \vspace{-4mm}
    \caption{Calculation of $p(c_t|E')$. As the code context to be completed is known, we leverage the LM to make predictions for each token within the context and determine if the predictions are accurate. These new observations can be used to calculate likelihood probability and to update the combined beliefs of the LM and retrieval module. Noted that for decoder-only models like GPT-3, this process does not incur additional computational costs.}
    \label{fig:likelihood}
    \vspace{-4mm}
\end{figure}

\subsection{Step 4. Combination of Inference Results}
\label{sec: combined_inference}
Combination of inference results from the LM and database is achievable by utilizing Equation~\ref{eq:total_probability}. Nevertheless, calculating $p(E'|c_t)$ directly poses challenges, as it represents the probability about the subsequent token of $c_t$, which remains unknown. We employ Bayes' rule to solve this, as:
\begin{equation}
    p(E'|c_t) \propto p(E') \times p(c_t|E')
\end{equation}

The prior probability, denoted as $p(E')$, has been computed using Equation~\ref{eq:prior_prob}. The likelihood probability $p(c_t|E')$ represents the probability of observing $c_t$ given the event $E'$. This likelihood probability indicates that the combination coefficient is also influenced by the code context being completed. Given that we already know $c_t$, for each token $x_i \in c_t$, we can predict the token $y_i$ by inputting the context $x_{0:i-1}$ into the language model. If $y_i = x_{i+1}$, we label it as event $E$, otherwise, it is classified as event $E'$. Bayesian analysis treats these newly occurring events as new observations, and they are utilized to update the prior probability.

In the context of parallel computing, our focus is on the latest $N-1$ observations within $c_t$ (can also be viewed as N-gram assumption). For illustration, as depicted in Fig.~\ref{fig:likelihood}, we utilize $y_{t-N-1:t-2}$ as the newly considered observations. Assuming that event $E'$ occurs $\alpha$ times, then event $E$ occurs $N-\alpha$ times. The computation of the likelihood probability $p(c_t|E')$ is expressed as follows:
\begin{equation}
    p(c_t|E') \approx p(x_{t-N-1:t-2}|E') = p(E')^{\alpha}(1-p(E'))^{N-\alpha}
\end{equation}
From the perspective of Bayesian inference, $p(E')$ is not a fixed number but a distribution represented by $\theta=p(E')$. Given that the distribution's expectation equals to $err$ (Equation~\ref{eq:prior_prob}), it is reasonable to consider $p(\theta) \sim \mathrm{Beta}(err \cdot N, (1-err) \cdot N)$\footnote{The choice of the Beta distribution is informed by its flexibility and suitability for modeling probabilities between 0 and 1~\cite{joyce2003bayes}. Additionally, we assume equal significance of prior beliefs and new observations, leading us to utilize the same $N$ for the Beta distribution.}. With the new evidence on $c_t$, we can deduce that the conditional distribution $p(\theta|c_t) \sim \mathrm{Beta}(err*N+\alpha, (1-err)*N+N-\alpha)$. As a result, the calculation of $p(y|c_t)$ can be formulated as:
\begin{equation}~\label{eq: bayesian_inference}
    \begin{aligned}
        p(y|c_t) &= \lambda p_{kNM}(y|c_t) + (1-\lambda)p_{LM}(y|c_t)\\
        \lambda &= p(E'|c_t) = \frac{1}{2}(\frac{\alpha}{N} + \frac{||\mathcal{M}||}{||\mathcal{D}||})
    \end{aligned}
\end{equation}
Equation~\ref{eq: bayesian_inference} incorporates two hyper-parameters: $k$ and $N$. In this context, $k$ is employed for the $k$-nearest neighbor search to compute $p_{kNM}(y|c_t)$, while $N$ represents the number of observations utilized for calculating the likelihood probability.

The values of $\alpha$, $||\mathcal{D}||$, and $||\mathcal{M}||$ in Equation~\ref{eq: bayesian_inference} are determined by three key components pertaining to the completion process: the code context to be completed, the specific language model utilized, and the chosen database. Specifically, when the error rate of the LM on the database and code context $c_t$ is elevated, $\lambda$ is correspondingly large, leading the combined approach to favor the prediction outcome of the retrieval module. In contrast, when the model's error rate on the database and code context $c_t$ is low, $\lambda$ is reduced, and the combination strategy tends to rely more on the prediction results generated by the LM.
\section{Experimental Setup}
\label{sec:experiment_setup}
To assess the efficacy of $k$NM-LM, we have formulated two research questions:
\begin{itemize}[leftmargin=*]
\item RQ1: What is the performance of $k$NM-LM for intra-project code completion?
\item RQ2: What is the performance of $k$NM-LM for intra-scenario code completion?
\end{itemize}

Furthermore, in relation to RQ1, we have explored the ability of $k$NM-LM to complete specific types of code. Regarding RQ2, we have investigated the effectiveness of $k$NM-LM in completing lines that involve Android API calls, along with an ablation study.
\subsection{Baselines}
As our proposed $k$NM-LM is a retrieval-augmented framework, for code completion, we choose two state-of-art pre-trained models as the base model: 
\begin{itemize}[leftmargin=*]
    \item CodeGPT~\cite{codexglue} is a GPT-based pre-trained model trained on the CodeSearchNet dataset~\cite{codesearchnet} for code completion and generation tasks. Specifically, we use CodeGPT-small-java-adaptedGPT2 for Java code and CodeGPT-small-py-adaptedGPT2 for Python code.
    \item UniXcoder~\cite{uniXcoder} is also pre-trained on the CodeSearchNet dataset and utilizes cross-modal content from code. It applies mask attention matrices with prefix adapters to control the model behavior. As UniXcoder is a cross-language model, we use UniXcoder-base for both Java and Python code.
\end{itemize}
We also compare $k$NM-LM with other competitive retrieval-augmented frameworks, including:
\begin{itemize}[leftmargin=*]
    \item \textbf{BM25~\cite{parvez2021retrieval}} is a block R-LM. It saves domain code as a document datastore, then searches the datastore to find code like the code to be completed based on the BM25~\cite{robertson2009probabilistic} algorithm. Finally, the search results are concatenated with the unfinished code and jointly fed into the language model.
    \item \textbf{ReACC}~\cite{ReACC} is also a block R-LM. Different from BM25, it uses GraphCodeBERT~\cite{graphcodebert} to train a retrieve model, and then encodes the domain code into a fixed length vector to store. ReACC only published their fine-tuned retrieval model on Python language, we only use it for Python language.
    \item \textbf{Hybrid} (BM25+ReACC)~\cite{ReACC} is an ensemble approach based on BM25 and ReACC. It combines search results from both BM25 and ReACC, then uses the re-ranked results to generate the next code token. Due to the need of using ReACC, we only use it for Python language.
    \item \textbf{$k$NN-LM}~\cite{knnlm} is a token R-LM, which saves each token in the domains code into a key-value datastore, and use $k$NN algorithm to retrieve similar code tokens. 
\end{itemize}

\subsection{Implementation Details}
In order to enable comparison with model fine-tuning (\cref{sec: compared_finetuning}), we employ the pre-trained Code-GPT and UniXcoder models as the base models, rather than relying on the black-box GPT-3 model. We utilize Elasticsearch for BM25 algorithm. ReACC is implemented using the released ``microsoft/reacc-py-retriever" as the retriever model. Hybrid uses the combination weight 0.9 from their original paper. In the case of $k$NN-LM and $k$NM-LM, we set $k$ (the number of retrieved nearest neighbors) to 8 for RQ1 and 1024 for RQ2 and employ L2 distance as the distance function for neighbor retrieval. We set $\lambda$ to 0.1 and window size $N$ to 8. In $k$NM-LM, we let the retrieval module to participate in computing the likelihood probability, if $y_i$ can be both correctly predicted by the LM and the retrieval module, we just ignore this token. Faiss-gpu~\cite{faiss} is used for accelerated $k$ nearest neighbors search, same as Khandelwal et al.~\cite{knnmt}.

\noindent \textbf{Metrices} Accuracy, representing the ratio of correctly predicting the next token, serves as the evaluation metric for token completion, while the metric for line completion involves Exact Match accuracy (EM) and Levenshtein Edit Similarity (ES)~\cite{svyatkovskiy2020intellicode}. To ensure fairness in the analysis of time and space complexity, all experiments are performed on a machine equipped with 2 NVIDIA 3090Ti-24GB GPU cards.

\begin{table}[t!]
\label{table:statistics_proj}
\centering
\caption{Statistics of intra-project completion dataset. $(\cdot)$ indicates the project's commit id. The test set contains methods created from the old commit to the latest commit.}
\begin{tabular}{l|cc}
\hline
Repo Name & Database & Test \\
\hline

Froyo\_Email &  2.4M (\href{https://github.com/Dustinmj/Froyo_Email/commit/56e48981f0093c737b38b7757410283be9ed823f}{56e48}) & 388K (\href{https://github.com/Dustinmj/Froyo_Email/commit/56e48981f0093c737b38b7757410283be9ed823f}{56e48} $\to$ \href{https://github.com/Dustinmj/Froyo_Email/commit/508c96313b0fe8bf7c0e0c4f069d0cd080f2d140}{508c9})\\

dropwizard & 2.2M (\href{https://github.com/dropwizard/dropwizard/commit/04c4563bd5dc6aeb8237ad5349cff05eb91fa5a8}{04c45}) & 388K (\href{https://github.com/dropwizard/dropwizard/commit/04c4563bd5dc6aeb8237ad5349cff05eb91fa5a8}{04c45} $\to$ \href{https://github.com/dropwizard/dropwizard/commit/ad30b6beaba15ca1cf4ef063aa6d128daba7af2d}{ad30b}) \\

AmazeFileManager & 1.9M (\href{https://github.com/TeamAmaze/AmazeFileManager/commit/d01c504cb55743c9b364110f8f714111a7b2b722}{d01c5}) & 292K (\href{https://github.com/TeamAmaze/AmazeFileManager/commit/d01c504cb55743c9b364110f8f714111a7b2b722}{d01c5} $\to$ \href{https://github.com/TeamAmaze/AmazeFileManager/commit/5c3cddf77ce80690da168457d7018edf43214c7a}{5c3cd}) \\

rest-assured & 1.8M (\href{https://github.com/rest-assured/rest-assured/commit/0c8f06a2f31ee40fa045cf4969ee0e2d262a3e47}{0c8f0}) & 274K (\href{https://github.com/rest-assured/rest-assured/commit/0c8f06a2f31ee40fa045cf4969ee0e2d262a3e47}{0c8f0} $\to$ \href{https://github.com/rest-assured/rest-assured/commit/246ba8740607e23f00097e04526a019685f22627}{246ba})\\

logging-log4j1 & 1.4M (\href{https://github.com/apache/logging-log4j1/commit/0d9e1787bc7b5f0d7932b8b5236d9e9a95692967}{0d9e1}) & 353K (\href{https://github.com/apache/logging-log4j1/commit/0d9e1787bc7b5f0d7932b8b5236d9e9a95692967}{0d9e1} $\to$ \href{https://github.com/apache/logging-log4j1/commit/e5c566e02eba058b8026f7fd3be96a7732b82a9d}{e5c56}) \\

feign & 1.3M (\href{https://github.com/OpenFeign/feign/commit/6989b503eff3e25a1036f8e8aabd0ec157f08a2c}{6989b}) & 140K (\href{https://github.com/OpenFeign/feign/commit/6989b503eff3e25a1036f8e8aabd0ec157f08a2c}{6989b} $\to$ \href{https://github.com/OpenFeign/feign/commit/d42fcef2bc9f470668e806f9215a09d7ddef0a5a}{d42fc}) \\

requery & 1.3M (\href{https://github.com/requery/requery/commit/a3f7105c124e5ce6d1ae7d1876061c6b7a97ca5f}{a3f71}) & 164K (\href{https://github.com/requery/requery/commit/a3f7105c124e5ce6d1ae7d1876061c6b7a97ca5f}{a3f71} $\to$ \href{https://github.com/requery/requery/commit/1d6fac733abf5a63a1bfcebd63b1e03674a9f9e3}{1d6fa}) \\

eureka & 1.2M (\href{https://github.com/Netflix/eureka/commit/225287ebb26b77ea68e15d744a31f8314f1f7425}{22528}) & 184K (\href{https://github.com/Netflix/eureka/commit/225287ebb26b77ea68e15d744a31f8314f1f7425}{22528} $\to$ \href{https://github.com/Netflix/eureka/commit/fc4e05a326cc7502b5b3ee041d8a8e4411ecfdf1}{fc4e0}) \\

galaxy & 1.2M (\href{https://github.com/galaxyproject/galaxy/commit/0ed700a7ea29035d9dbab01363a5b4ed67aee2e7}{0ed70}) & 166K (\href{https://github.com/galaxyproject/galaxy/commit/0ed700a7ea29035d9dbab01363a5b4ed67aee2e7}{0ed70} $\to$ \href{https://github.com/galaxyproject/galaxy/commit/8fcc63f8c7a1f3bf5ac02d68c7fcc3d474315030}{8fcc6}) \\

interview & 1.1M (\href{https://github.com/mission-peace/interview/commit/ed0d5ef4daf330999cc50c62d3f639b4e44846e4}{ed0d5})  & 164K (\href{https://github.com/mission-peace/interview/commit/ed0d5ef4daf330999cc50c62d3f639b4e44846e4}{ed0d5} $\to$ \href{https://github.com/mission-peace/interview/commit/94be5deb0c0df30ade2a569cf3056b7cc1e012f4}{94be5}) \\
\hline

android-priority-jobqueue & 592K (\href{https://github.com/yigit/android-priority-jobqueue/commit/0287d52e38e3b05d6fa5ec6ae11cd8b7635eafff}{0287d})  & 69K (\href{https://github.com/yigit/android-priority-jobqueue/commit/0287d52e38e3b05d6fa5ec6ae11cd8b7635eafff}{0287d} $\to$ \href{https://github.com/yigit/android-priority-jobqueue/commit/f807cdd4bc8425944702182fa66de59b36193c00}{f807c})\\

lottie-android & 541K (\href{https://github.com/airbnb/lottie-android/commit/c32f18ede7873f4c5e16493456340ef9878830d9}{c32f1}) & 88K (\href{https://github.com/airbnb/lottie-android/commit/c32f18ede7873f4c5e16493456340ef9878830d9}{c32f1} $\to$ \href{https://github.com/airbnb/lottie-android/commit/42b486156b654bace262d3437ae9d775804e3811}{42b48}) \\

xUtils3 & 472K (\href{https://github.com/wyouflf/xUtils3/commit/0d10d68a67218dac7b14687dcf7ea8122d32b831}{0d10d}) & 35K (\href{https://github.com/wyouflf/xUtils3/commit/0d10d68a67218dac7b14687dcf7ea8122d32b831}{0d10d} $\to$ \href{https://github.com/wyouflf/xUtils3/commit/4f3c36f62249a44a2d08942224b6cc8dcd7f7ed8}{4f3c3}) \\

Fragmentation & 408K (\href{https://github.com/YoKeyword/Fragmentation/commit/0e972b841adbf53c0fb76d4fa38cd8a8b0e0f5ee}{0e972}) & 59K (\href{https://github.com/YoKeyword/Fragmentation/commit/0e972b841adbf53c0fb76d4fa38cd8a8b0e0f5ee}{0e972} $\to$ \href{https://github.com/YoKeyword/Fragmentation/commit/d5e79a3d953ec688006a7a96f460b7d732d4c441}{d5e79}) \\

VirtualAPK & 308K (\href{https://github.com/didi/VirtualAPK/commit/e6174132a0e83c49ed8f70561b5a94c59951b195}{e6174})  & 15K (\href{https://github.com/didi/VirtualAPK/commit/e6174132a0e83c49ed8f70561b5a94c59951b195}{e6174} $\to$ \href{https://github.com/didi/VirtualAPK/commit/01b73298b0b00ce6a22a27be5a95f56b99a13ada}{01b73}) \\

spring-boot-starter & 181K (\href{https://github.com/mybatis/spring-boot-starter/commit/f0951566ecc21f3166687b691665de994534b196}{f0951}) & 16K (\href{https://github.com/mybatis/spring-boot-starter/commit/f0951566ecc21f3166687b691665de994534b196}{f0951} $\to$ \href{https://github.com/mybatis/spring-boot-starter/commit/a634e82e6a107a681cf96a9f24f5c75cee5a1eca}{a634e}) \\

DiscreteScrollView & 147K (\href{https://github.com/yarolegovich/DiscreteScrollView/commit/4967100d81b468f50febc1a901713847f8d5cd0f}{49671}) & 15K (\href{https://github.com/yarolegovich/DiscreteScrollView/commit/4967100d81b468f50febc1a901713847f8d5cd0f}{49671} $\to$ \href{https://github.com/yarolegovich/DiscreteScrollView/commit/9d979827f29b4f5e2b1c88cddb5c38ee98e16546}{9d979})\\

flow & 108K (\href{https://github.com/facebook/flow/commit/489b2d02c07b76b4e4aa8cb60a625b7f24ebcb35}{489b2}) & 18K (\href{https://github.com/facebook/flow/commit/489b2d02c07b76b4e4aa8cb60a625b7f24ebcb35}{489b2} $\to$ \href{https://github.com/facebook/flow/commit/06eb09503b34d94a7922a9c1189508e99aa5035f}{06eb0}) \\

hover & 259K (\href{https://github.com/google/hover/commit/7e95cf85248eb6e4f2cc40922cb6e34adedb830c}{7e95c}) & 14K (\href{https://github.com/google/hover/commit/7e95cf85248eb6e4f2cc40922cb6e34adedb830c}{7e95c} $\to$ \href{https://github.com/google/hover/commit/03e448dea7dab033da4890960eec415a2b4cf07b}{03e44})\\

StickyHeaderListView & 117K (\href{https://github.com/sunfusheng/StickyHeaderListView/commit/002657beee840f144419e6744e861c997007107f}{00265}) & 6K (\href{https://github.com/sunfusheng/StickyHeaderListView/commit/002657beee840f144419e6744e861c997007107f}{00265} $\to$ \href{https://github.com/sunfusheng/StickyHeaderListView/commit/b871cb53e88410134cc03b9b5076c4b95f4a3a28}{b871c}) \\

\hline
\end{tabular}%
\label{tab:my_label}
\end{table}

\section{RQ1: Intra-Project Code Completion}
\label{sec:rq1}
\subsection{Study Design}
Auto code completion plugins are widely used in projects to streamline the coding process, with other files in the project often containing helpful information for code completion. Consequently, we have developed the intra-project code completion task. In this research question, we utilize Git commit history to determine the creation time of various methods in the project. We then select the project snapshot at a specific commit as the basis for the written code used to build the database. Methods created after that commit serve as a test set to evaluate the performance of different approaches.

\begin{table*}[t!]
  \centering
  \caption{Token completion Accuracy of different code types on intra-project dataset. Numbers are shown in percentage(\%).}
  \resizebox{\linewidth}{!}{%
\begin{tabular}{c|lr|cccc|cccc}
\hline
      & Types & Percent & \cellcolor[rgb]{ .906,  .902,  .902}CodeGPT & +BM25  & +$k$NN-LM & \cellcolor[rgb]{ .741,  .843,  .933}+$k$NM-LM & \cellcolor[rgb]{ .906,  .902,  .902}UniXcoder & +BM25  & +$k$NN-LM & \cellcolor[rgb]{ .741,  .843,  .933}+$k$NM-LM\\
\hline
\multirow{6}[2]{*}{Large Projects} & All   & 100.00  & \cellcolor[rgb]{ .906,  .902,  .902}54.59  & 56.12  & 58.69  & \cellcolor[rgb]{ .741,  .843,  .933}\textbf{69.66}  & \cellcolor[rgb]{ .906,  .902,  .902}53.11  & 59.16  & 57.46  & \cellcolor[rgb]{ .741,  .843,  .933}\textbf{68.96} \\
      &Punctuation & 34.56  & \cellcolor[rgb]{ .906,  .902,  .902}64.97  & 66.74  & 68.10  & \cellcolor[rgb]{ .741,  .843,  .933}\textbf{86.07}  & \cellcolor[rgb]{ .906,  .902,  .902}67.24  & 73.40  & 70.70  & \cellcolor[rgb]{ .741,  .843,  .933}\textbf{84.90}  \\
      &Identifier & 30.63  & \cellcolor[rgb]{ .906,  .902,  .902}38.37  & 40.72  & 41.48  & \cellcolor[rgb]{ .741,  .843,  .933}\textbf{47.68}  & \cellcolor[rgb]{ .906,  .902,  .902}38.43  & 46.69  & 41.42  & \cellcolor[rgb]{ .741,  .843,  .933}\textbf{47.33 } \\
      &Operator & 19.51  & \cellcolor[rgb]{ .906,  .902,  .902}70.02  & 70.66  & 71.39  & \cellcolor[rgb]{ .741,  .843,  .933}\textbf{77.05}  & \cellcolor[rgb]{ .906,  .902,  .902}64.84  & 69.74  & 66.29  & \cellcolor[rgb]{ .741,  .843,  .933}\textbf{81.08}  \\
      &Keyword & 8.02  & \cellcolor[rgb]{ .906,  .902,  .902}49.96  & 51.17  & 53.82  & \cellcolor[rgb]{ .741,  .843,  .933}\textbf{56.97}  & \cellcolor[rgb]{ .906,  .902,  .902}28.96  & 34.46  & 32.21  & \cellcolor[rgb]{ .741,  .843,  .933}\textbf{46.51}  \\
      &Literals & 7.28  & \cellcolor[rgb]{ .906,  .902,  .902}39.87  & 38.67  & 59.08  & \cellcolor[rgb]{ .741,  .843,  .933}\textbf{76.54}  & \cellcolor[rgb]{ .906,  .902,  .902}41.36  & 42.01  & 63.87  & \cellcolor[rgb]{ .741,  .843,  .933}\textbf{75.86} \\
\hline
\multirow{6}[2]{*}{Small Projects} & All   & 100.00  & \cellcolor[rgb]{ .906,  .902,  .902}53.71  & 55.10  & 56.59  & \cellcolor[rgb]{ .741,  .843,  .933}\textbf{66.70}  & \cellcolor[rgb]{ .906,  .902,  .902}51.00  & 56.92  & 54.05  & \cellcolor[rgb]{ .741,  .843,  .933}\textbf{64.19}\\
      & Punctuation & 37.30  & \cellcolor[rgb]{ .906,  .902,  .902}63.86  & 65.39  & 67.38  & \cellcolor[rgb]{ .741,  .843,  .933}\textbf{85.57}  & \cellcolor[rgb]{ .906,  .902,  .902}64.77  & 71.70  & 68.01  & \cellcolor[rgb]{ .741,  .843,  .933}\textbf{85.05}  \\
      & Identifier & 32.06  & \cellcolor[rgb]{ .906,  .902,  .902}37.76  & 39.11  & 40.29  & \cellcolor[rgb]{ .741,  .843,  .933}\textbf{44.48}  & \cellcolor[rgb]{ .906,  .902,  .902}35.96  & 42.49  & 38.47  & \cellcolor[rgb]{ .741,  .843,  .933}\textbf{40.10}  \\
      & Operator & 17.54  & \cellcolor[rgb]{ .906,  .902,  .902}69.83  & 70.86  & 70.46  & \cellcolor[rgb]{ .741,  .843,  .933}\textbf{73.48}  &\cellcolor[rgb]{ .906,  .902,  .902} 63.05  & 66.79  & 64.41  & \cellcolor[rgb]{ .741,  .843,  .933}\textbf{76.12}  \\
      & Keyword & 10.55  & \cellcolor[rgb]{ .906,  .902,  .902}48.92  & 50.12  & 50.00  & \cellcolor[rgb]{ .741,  .843,  .933}\textbf{55.62}  & \cellcolor[rgb]{ .906,  .902,  .902}32.01  & 37.19  & 35.09  & \cellcolor[rgb]{ .741,  .843,  .933}\textbf{43.19}  \\
      & Literals & 2.56  & \cellcolor[rgb]{ .906,  .902,  .902}27.44  & 29.84  & 48.26  & \cellcolor[rgb]{ .741,  .843,  .933}\textbf{74.59}  & \cellcolor[rgb]{ .906,  .902,  .902}30.09  & 30.96  & 45.87  & \cellcolor[rgb]{ .741,  .843,  .933}\textbf{72.29}\\
\hline
\end{tabular}%
    }
  \label{tab:proj_acc}%
\end{table*}%

\subsection{Dataset}
We compile an intra-project code completion dataset by collecting 10 large projects (over 1M in size) and 10 small projects (under 1M in size) from the test set of Java-mid~\cite{java_mid}. The dataset's statistics are displayed in Table~\ref{table:statistics_proj}. Following the approach of Bogomolov et al.\cite{project_miner}, we gather the Git history for each project and extract all method creation commits using Miner\cite{psiminer}. These methods are sorted by commit time, and we select the commit at which the first 80\% of methods have been created. The project snapshot at this commit is used to build the database, while the remaining 20\% of methods serve as the test set. Then, we separately evaluate the performance on each project.
\begin{figure}[!t]
    \centering
    \includegraphics[width=\linewidth]{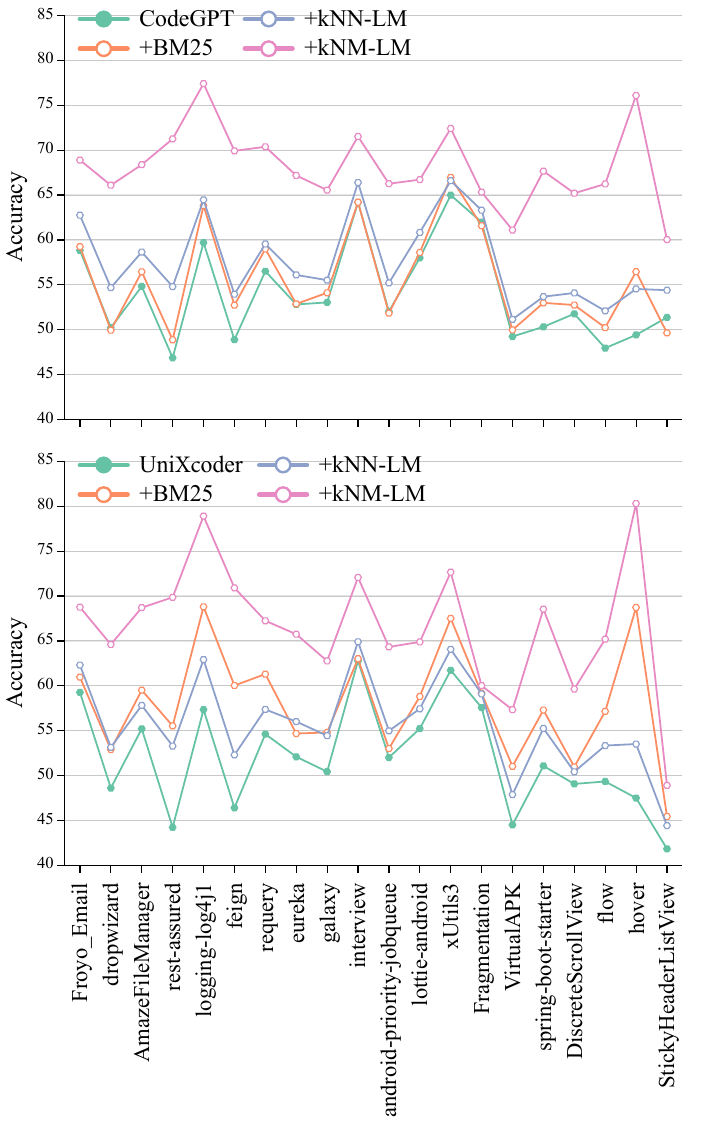}
    \caption{Token completion accuracy on intra-project dataset. The abscissa is the project name, ordered from large to small based on the project size.}
    \label{fig:acc_in_each_proj}
    \vspace{-4mm}
\end{figure}
\begin{table}[t!]
  \centering
  \caption{Statistics of Intra-scenario Completion Dataset (API-Bench).}
    \begin{tabular}{c|c|cc|cc}
    \hline
    \multirow{2}[4]{*}{PL} & \multirow{2}[4]{*}{Domain} & \multicolumn{2}{c|}{\#num of projects} & \multicolumn{2}{c}{\#num of functions} \\
\cline{3-6}          &       & Database & Test  & Database & Test \\
    \hline
    \multirow{4}[2]{*}{Java} & Android & 292   & 85    & 53,646 & 29,211\\
          & ML    & 38    & 14    & 27,962 & 10,435 \\
          & Security & 42    & 16    & 14,202 & 4,698 \\
          & Test  & 41    & 14    & 16,439 & 5,341 \\
    \hline
    \multirow{4}[2]{*}{Python} & DL    & 191   & 116   & 24,546 & 12,778 \\
          & ML    & 165   & 159   & 23,990 & 20,533 \\
          & Security & 95    & 47    & 10,823 & 4,527 \\
          & Test  & 57    & 22    & 4,761 & 3,303 \\
    \hline
    \end{tabular}%
    \label{tab:statics_api_bench}%
\end{table}%

\begin{table*}[t!]
  \centering
  \caption{Token completion accuracy on intra-scenario dataset (API-Bench). Numbers are shown in percentage(\%). $k$NM-LM w/o Bayesian uses the manually set $\lambda$ like $k$NN-LM~\cite{knnlm}, and $k$NM-LM w/o new observations uses $p(E')$ as $\lambda$.}
\begin{tabular}{l|cccc|cccc}
\hline
\multicolumn{1}{c|}{\multirow{2}[2]{*}{Models}} & \multicolumn{4}{c|}{Java}     & \multicolumn{4}{c}{Python} \\
      & Android & ML    & Security & Test  & DL    & ML    & Security & Test \\
\hline
\rowcolor[rgb]{ .906,  .902,  .902}CodeGPT & 66.77 & 64.98 & 65.4  & 64.74 & 51.71 & 51.81 & 52.77 & 50.09 \\
+BM25  & 66.75 & 64.28 & 63.92 & 64.08 & 49.96 & 49.94 & 50.27 & 47.14 \\
+ReACC & -     & -     & -     & -     & 49.92 & 49.98 & 50.45 & 47.71 \\
+Hybrid & -     & -     & -     & -     & 52.16 & 52.32 & 52.94 & 50.37 \\
+$k$NN   & 69.88 & 67.68 & 68.14 & 67.12 & 53.65 & 53.76 & 56.83 & 53.98 \\
\rowcolor[rgb]{ .741,  .843,  .933} +$k$NM   & \textbf{72.34} & \textbf{71.00}    & \textbf{71.40}  & \textbf{70.54} & \textbf{59.50}  & \textbf{59.27} & \textbf{64.85} & \textbf{60.87} \\
\rowcolor[rgb]{ .741,  .843,  .933} +$k$NM w/o Bayesian & 69.38 & 67.45 & 67.86 & 66.90  & 53.46 & 53.58 & 56.59 & 53.66 \\
\rowcolor[rgb]{ .741,  .843,  .933} +$k$NM w/o new observations & 72.17 & 70.90  & 71.30  & 70.36 & 59.49 & 59.26 & 64.68 & 60.73\\
\hline
\rowcolor[rgb]{ .906,  .902,  .902}UniXcoder & 67.75 & 67.65 & 66.80  & 65.65 & 59.53 & 59.77 & 63.01 & 60.30\\
+BM25  & 67.59 & 66.51 & 65.60  & 64.88 & 56.99 & 56.66 & 61.49 & 56.90 \\
+ReACC & -     & -     & -     & -     & 56.30  & 56.29 & 59.32 & 56.53 \\
+Hybrid & -     & -     & -     & -     & 56.10  & 56.41 & 59.76 & 57.20 \\
+$k$NN   & 69.83 & 71.28 & 69.20  & 67.63 & 62.30  & 62.59 & 69.86 & 65.88 \\
\rowcolor[rgb]{ .741,  .843,  .933} +$k$NM   & \textbf{74.86} & 76.08 & \textbf{74.84} & \textbf{73.12} & 67.03 & 67.02 & \textbf{75.56} & \textbf{70.09}\\
\rowcolor[rgb]{ .741,  .843,  .933} +$k$NM w/o Bayesian & 69.76 & 71.22 & 69.15 & 67.55 & 62.24 & 62.55 & 69.88 & 65.84 \\
\rowcolor[rgb]{ .741,  .843,  .933} +$k$NM w/o new observations & 74.83 & \textbf{76.10}  & 74.70  & 73.07 & \textbf{67.07} & \textbf{67.07} & 75.56 & 70.09 \\
\hline
\end{tabular}%
\label{tab:acc_in_scenario}%
\end{table*}%

\subsection{Results}
\noindent \textbf{Token Completion Accuracy.} 
As illustrated in Fig.~\ref{fig:acc_in_each_proj}, our proposed approach, $k$NM-LM, exhibits a significant improvement for all projects in the dataset. $k$NM-LM achieves the most substantial improvement on the ``hover" project, with a 26.67\% and 35.99\% increase with CodeGPT and UniXcoder, respectively. The trend of the curve indicates that larger projects benefit more from using $k$NM-LM. This may be because large projects contain more code in the database, making it more likely to find useful information from the database. We also observe that the performance boost gained by using the prompt method (BM25) is inconsistent across different projects. In some cases, it may be worse than the base model, such as with CodeGPT on the ``android-priority-jobqueue" project, where it decreases from 52.03\% to 51.84\%. This discrepancy may arise due to the absence of code in the database that closely resembles the code requiring completion. When we merge the retrieved result with the input and feed them into the language model, there is a potential to introduce irrelevant noise that does not contribute to the accurate completion.

\noindent \textbf{Analysis of Completing Different Code Token Types.} 
Language models often face greater difficulties when dealing with certain code token types. For instance, predicting identifiers (such as variable or function names) is more challenging than predicting keywords (like ``if" or ``for"). However, developers primarily utilize code completion plugins to finalize intra-project APIs~\cite{hellendoorn2019code}. Hence, we follow previous works~\cite{izadi2022codefill} to investigate the performance of models when predicting different types of tokens. Results are presented in Table~\ref{tab:proj_acc}. From the results, we observe that all retrieval-augmented models can enhance the code completion performance of the base model in intra-project contexts, with $k$NM-LM showing the most significant improvement. Specifically, for completing punctuation, which is related to programming language syntax, $k$NM-LM improves the base models with increases ranging from 17.66\% to 21.71\%. For completing identifiers and literals, which are project-specific components, the base models are subpar. This supports our assumption that general code completion language models lack domain-specific knowledge. Using $k$NM-LM, the accuracy of completing identifiers exhibited improvements of 9.31\% and 8.90\% in large projects, and 6.72\% and 4.14\% in small projects, respectively. This notable enhancement can be attributed to $k$NM-LM's use of project code as the retrieval database, allowing domain-specific knowledge, such as custom functions and objects within the project, to be incorporated into the LM.
\begin{tcolorbox}[colframe=gray!40!gray,left = 0mm,right=0mm,top = 0mm,bottom=0mm]
\textbf{$\blacktriangleright$RQ1$\blacktriangleleft$} Results indicate that utilizing historical project code with R-LMs improves intra-project code completion, with $k$NM-LM exhibiting the most substantial improvement across all projects compared with baselines.
\end{tcolorbox}

\section{RQ2: Intra-Scenario Code Completion}
\label{sec:rq2}

\subsection{Study Design}
Completing code across different scenarios necessitates specialized knowledge, primarily manifested in understanding and utilizing the specific third-party libraries. For instance, mobile development code uses the Android library, while deep learning code depends on the PyTorch library. However, large language models may lack insight into the usage of these third-party libraries, due to their proprietary or frequently updated nature. In this research, we leverage projects that use the same third-party libraries to build the code scenario database. These projects contain scenario-specific knowledge, including distinct API usage patterns. Then, we explore the potential advantages of using this scenario database for intra-scenario code completion. Furthermore, we evaluate the effectiveness of completing lines of code that involve Android API calls.
 
\subsection{Dataset}
We utilize the API-Bench dataset~\cite{api_bench} to facilitate code completion within specific scenarios. This dataset encompasses four distinct code scenarios, involving both Java and Python. In each of these scenarios, we employ the train set to construct a datastore and then evaluate token completion using the corresponding test set. For a comprehensive overview of the dataset's statistics, refer to Table~\ref{tab:statics_api_bench}.

In addition, we curate an extra test set designed to evaluate the effectiveness of completing lines containing Android API calls. To achieve this, we leverage JavaParser\footnote{\url{https://github.com/javaparser/javaparser}} to extract the object type associated with the method call, filtering out API calls from the Android library. Subsequently, we utilize the code context preceding the API call to facilitate completion of the entire code line, encompassing the API name and input parameters. This dedicated test set comprises 49,344 lines featuring Android API usage. Notably, the database employed for this task is identical to the one used for token completion within the Android dataset.
\subsection{Results}
\noindent \textbf{Token Completion Accuracy.}
Results in Table~\ref{tab:acc_in_scenario} suggest that R-LMs do not consistently outperform the base LMs in intra-scenario completion task. For example, in Java, BM25 has a negative impact on Android, ML, and Security. Similarly, in Python, both BM25 and ReACC perform worse than the base LMs across all four scenarios. While the Hybrid and $k$NN-LM models show some improvement, their performance gains are smaller than those observed in the intra-project completion. This is mainly because code within the same project is more similar than those in the same scenario~\cite{structuralLocality}, making it harder for the retrieval-augmented approaches to find useful code. On the other hand, $k$NM-LM shows a significant performance improvement of approximately 6-12\% compared to the base LMs. This improvement can be attributed to the decoupled database construction. Traditional R-LMs retain all data when constructing the database, leading to the code that use the specific API call not being retrieved due to the few samples problem. In contrast, $k$NM-LM constructs a decoupled database that only retains code that the LM cannot predict accurately. This strategy increases the ratio of those specific API calls that cannot be correctly predicted by LM in the datastore, thus improving performance.

\noindent \textbf{Ablation Study of $k$NM-LM.}
Table \ref{tab:acc_in_scenario} also includes an ablation study on the use of Bayesian inference. Specifically, two variants of the $k$NM-LM are evaluated: $k$NM-LM w/o Bayesian, which relies on a manually set $\lambda$ value, and $k$NM-LM w/o new observations, which sets the prior probability of $p(E')$ as $\lambda$. The results indicate that manually setting the combination coefficient has the least favorable effect, while employing Bayesian inference has the most favorable effect in most cases. This is primarily because determining the optimal $\lambda$ value for different scenarios and LMs can be challenging, while Bayesian inference can take into account the impact of both the scenarios and LMs and dynamically adjust $\lambda$ based on the code text being completed.

\noindent \textbf{Performance of Completing Lines that involve Android APIs.}
Table~\ref{tab:android_line_com} presents an experimental evaluation of completing lines containing Android API. The results show that the performance of the basic LM is not satisfactory, as indicated by the EM metrics of 5.99\% and 2.74\% for completion of lines that contain Android API calls. However, the EM metrics of $k$NM-LM are significantly improved, increasing by 7.75\% and 8.94\%, respectively. Moreover, we utilize the Venn diagram~\cite{bardou2014jvenn} to examine the precision of each model's completion on a per-method basis. The analysis demonstrates that $k$NM-LM outperforms other models in terms of accuracy, with 4,245 more accurately predicted code lines than all other models combined.

\begin{tcolorbox}[colframe=gray!40!gray,left = 0mm,right=0mm,top = 0mm,bottom=0mm]
\textbf{$\blacktriangleright$RQ2$\blacktriangleleft$} Utilizing code from the same scenario can improve intra-scenario code completion, with $k$NM-LM exhibits maximum enhancement. Also, Bayesian inference is important in determining the optimal combination coefficient. 
\end{tcolorbox}

\section{Discussion}
\begin{table}[t!]
  \centering
  \caption{Performance of completing lines that contain Android APIs.}
    \begin{tabular}{l|cc|cc}
    \hline
    \multirow{2}[4]{*}{} & \multicolumn{2}{c|}{CodeGPT} & \multicolumn{2}{c}{UniXcoder}\\
\cline{2-5}          & ES    & EM    & ES    & EM\\
    \hline
    \rowcolor[rgb]{ .906,  .902,  .902}Base  & 44.38  & 5.99  & 42.87  & 2.74\\
    +BM25  & 44.65  & 6.02 & 42.77 & 2.71  \\
    +$k$NN-LM & 48.48  & 7.91 & 44.79  & 3.40\\
    \rowcolor[rgb]{ .741,  .843,  .933} +$k$NM-LM & \textbf{54.10}  & \textbf{13.74} & \textbf{49.59}  & \textbf{11.68}\\
    \hline
    \end{tabular}%
  \label{tab:android_line_com}%
\end{table}%

In this chapter, our primary focus is on assessing the effectiveness of $k$NM-LM from three key angles. As $k$NM-LM is an improvement over $k$NN-LM, we begin by discussing the distinctions between these two methods. Following this, since the aim of $k$NM-LM is to enable domain-specific code completion without the need of fine-tuning, we proceed to compare it with fine-tuned models. Lastly, we conduct case studies to provide concrete examples that validate the efficacy of our proposed approach.

\begin{figure}[!t] 
\centering 
\includegraphics[width=0.7\linewidth]{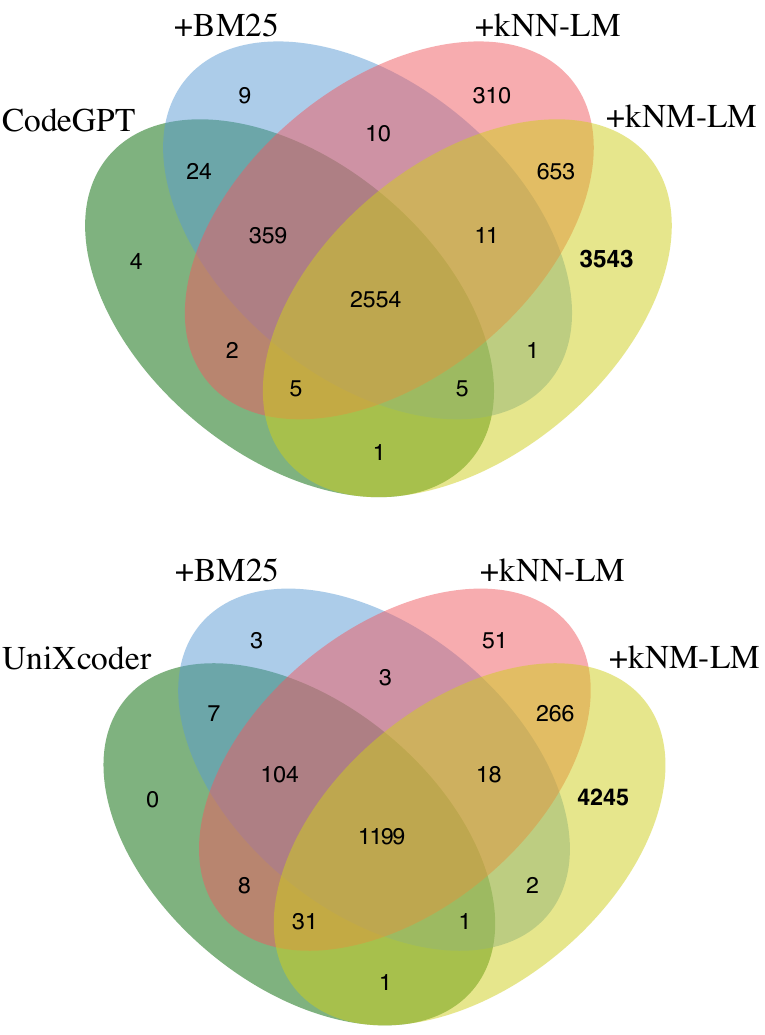} \caption{Venn diagram of completing lines that contain Android APIs. It shows the number of samples that are completed correctly.} 
\label{fig:vnn_diagram}
\end{figure} 

\begin{table}[t]
  \centering
  \caption{Comparison of resource usage in $k$NN-LM and $k$NM-LM.}
    \begin{tabular}{l|cccc}
    \hline
    \multirow{3}[4]{*}{Model} & \multicolumn{2}{c}{Large projects} & \multicolumn{2}{c}{Small Projects}\\
\cline{2-5}          & Speed & DB size  & Speed  & DB size\\
          & (tokens/s) & (MB)  & (tokens/s) & (MB)\\
    \hline
    \rowcolor[rgb]{ .906,  .902,  .902}CodeGPT & 3063.40  & -  & 1773.92  & -\\
    +$k$NN-LM   & 1242.25  & 780.48  & 990.71  & 156.78  \\
    \rowcolor[rgb]{ .741,  .843,  .933}+kNM-LM   & 1282.68  & 391.12  & 953.27  & 79.17\\
    \hline
    \rowcolor[rgb]{ .906,  .902,  .902}UniXcoder & 2730.87  & -  & 1625.89  & -\\
    +$k$NN-LM   & 1175.03  & 713.76  & 937.06  & 142.42  \\
    \rowcolor[rgb]{ .741,  .843,  .933}+$k$NM-LM   & 1252.48  & 428.96  & 948.49  & 86.85\\
    \hline
    \end{tabular}%
  \label{tab:resource_use}%
\end{table}%

\subsection{Compared with kNN-LM.}
\noindent \textbf{Speed and Space Usage Analysis.}
Table~\ref{tab:resource_use} demonstrates that both $k$NN-LM and $k$NM-LM have a negative impact on completion speed. However, even with the worst-case scenario, the completion of 937.06 words can be achieved within 1 second. As developers typically anticipate receiving the completion results within 200 milliseconds~\cite{wang2023practitioners}, we think that the impact on completion speed is not significant. Storage space consumption, on the other hand, is a more significant consideration. Both models require database storage, but $k$NM-LM uses considerably less space than $k$NN-LM. In large projects, $k$NM-LM saves 284.83MB and 389.36MB compared to $k$NN-LM, while in small projects, it saves 55.57MB and 67.61MB. Given the significant performance gain from $k$NM-LM, the additional storage space required is acceptable.

\begin{figure}[!t]
    \centering
    \includegraphics[width=\linewidth]{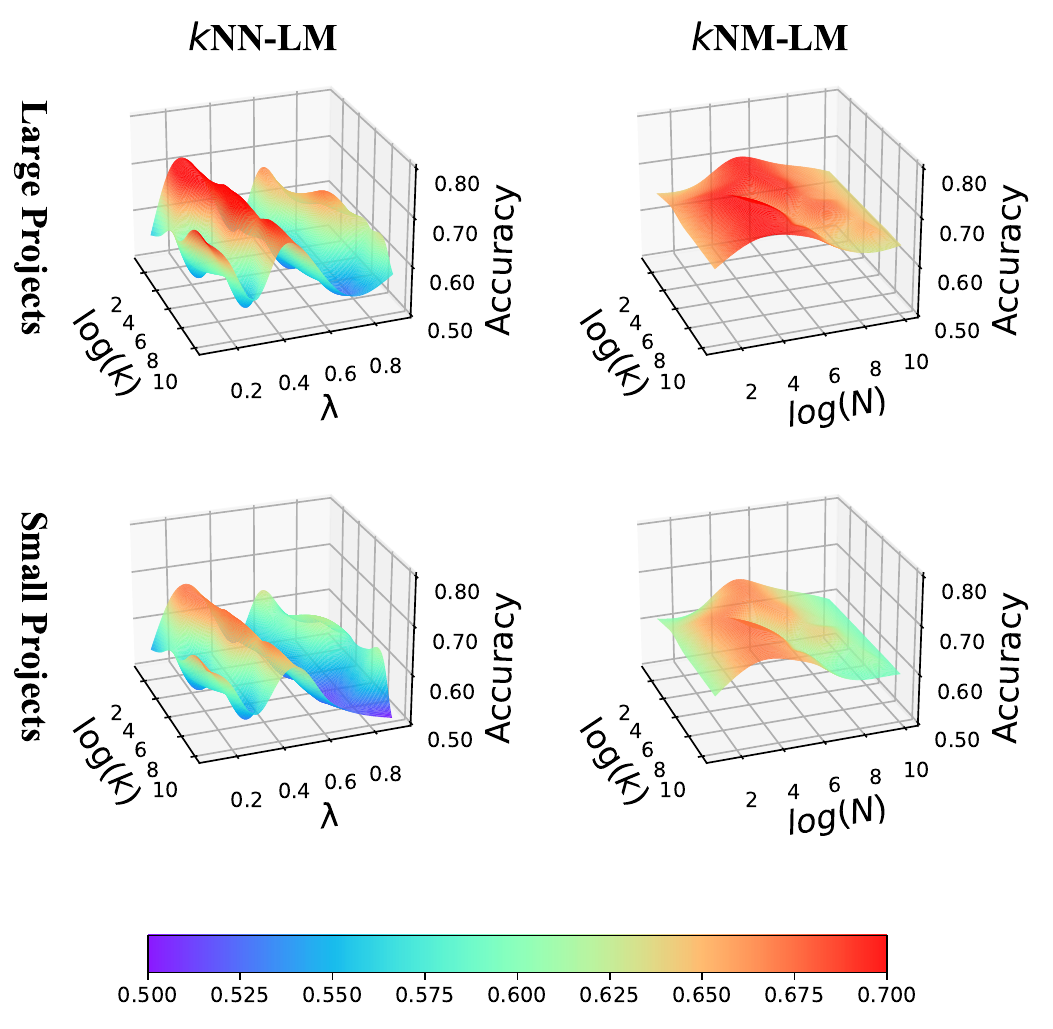}
    \caption{Influence of hyper-parameters settings in $k$NN-LM and $k$NM-LM. The base LM is CodeGPT and the task is intra-project code completion. The number of searched neighbors $k$ is used in both. $k$NN-LM needs to set the combination coefficient $\lambda$, and $k$NM-LM needs to set the window size $N$. The above two sub-figures are the results on large projects, and the below two are results on small projects.}
    \label{fig:hypeparams}
\end{figure}

\noindent \textbf{Influence of Hyper-parameters.}
We compare the performance of $k$NM-LM and $k$NN-LM with varying hyper-parameter settings for intra-project code completion. Results, shown in Fig.~\ref{fig:hypeparams}, indicate that hyper-parameters have a significant impact on the performance of $k$NN-LM. In contrast, $k$NM-LM is less sensitive to hyper-parameters and achieves good performance for most settings. Therefore, $k$NM-LM is more suitable for domain adaptive code completion task, as it can provide satisfactory results without the carefully hyper-parameters configuration.

\subsection{Compared with Fine-tuning.}
\label{sec: compared_finetuning}
\begin{figure}[!t]
    \centering
    \includegraphics[width=\linewidth]{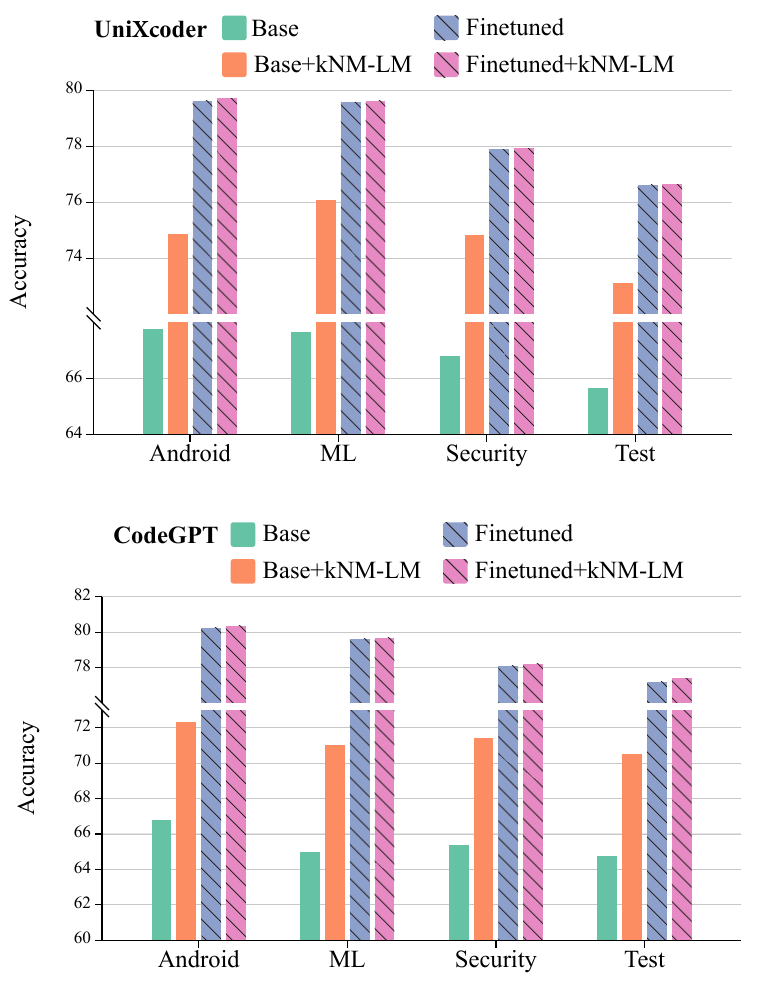}
    \caption{Finetune VS $k$NM-LM.}
    \label{fig:finetune_vs_knm_lm}
    \vspace{-6mm}
\end{figure}

We compared $k$NM-LM and fine-tuning on intra-scenario code completion task, as presented in Fig.~\ref{fig:finetune_vs_knm_lm}. We found that fine-tuning is more effective than $k$NM-LM, but using $k$NM-LM after fine-tuning can still lead to some improvement. It is worth noting that fine-tuning may not be a practical solution for domain adaptive code completion due to the continuously updated domain features and the black-box constraint of LM. In contrast, $k$NM-LM may be more suitable for individual users to obtain personalized code completion service.

\subsection{Qualitative Analysis.}
\begin{figure}[!t]
    \centering
    \includegraphics[width=\linewidth]{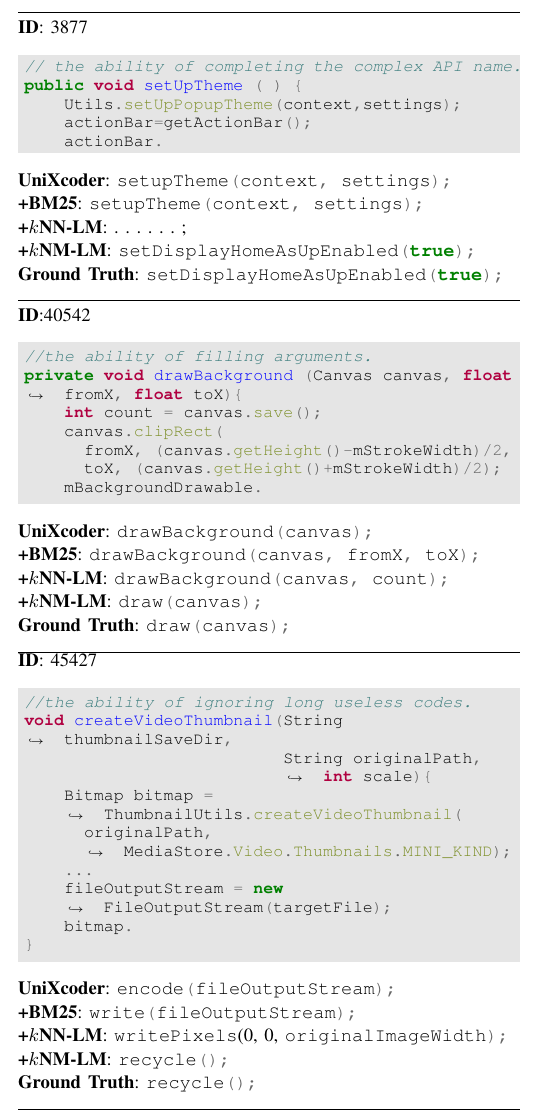}
    \caption{\small Qualitative examples.}
    \label{fig:qualitative_examples}
    \vspace{-6mm}
\end{figure}

In this study, we conducted a qualitative analysis on completing lines that contain Android API calls, presented in Fig.~\ref{fig:qualitative_examples}.
The study also included three examples that demonstrated the capabilities of the models used. For instance, Example ID=3877 showed that $k$NM-LM accurately completed complex API names, which a general model without Android-specific knowledge would fail to predict accurately. Example ID=40542 demonstrated that $k$NM-LM accurately predicted the API name and correctly filled in the corresponding input parameters. On the other hand, the BM25 and $k$NN-LM models failed to correct the API name and also incorrectly completed additional input parameters. Lastly, example ID=45427 verified that $k$NM-LM can ignore irrelevant code and accurately complete the intended code. In contrast, other models were affected by irrelevant code, including file writing semantics in the completed code.

\section{THREATS TO VALIDITY}
The first threat is related to the base LMs selected for the experiment. To mitigate this threat, we chose two state-of-the-art pretrained LMs, CodeGPT and UniXcoder, for the code completion task.  
We want to compare our proposed method with fine-tuning, hence we do not use black-box LM systems, such as Codex or ChatGPT. However, in~\cref{sec: inference_lm}, we introduced how to use our approach in GPT-3 through API calling. The second threat is related to the setting of the domain database and test set on intra-project code completion, as both use code from the same project. To mitigate this threat, we use the Git commit history to split the code for building the database and the code for testing, ensuring that the code in the test set were developed after those in the database. The third threat is related to the fact that we did not verify the effect of ReACC and Hybrid on the Java language, as the authors only released their pretrained models on Python. To mitigate this threat, we chose to compare our approach with BM25, which has similar performance compared to ReACC and Hybrid as reported in their paper. Lastly, in the context of intra-project code completion, our choice of the commit point at which 80\% of the code is already written introduces a potential concern. This choice does not always reflect practical scenarios, we intend to investigate the impact of varying the size of the intra-project database in our future work. To ensure transparency and enable verification by other researchers, we open-sourced all our code, data, and experimental results.
\section{Related work}

\subsection{Code Completion}
Code completion can be divided into three categories based on the code type to be completed: API name completion~\cite{he2021pyart,pythia}, variable name completion~\cite{li2018codewithPointer,allamanCode2Graph} and arbitrary token completion~\cite{izadi2022codefill,guo2021learning,retrieveNeuralCodeGen,retrieveAndEdit}. API name completion aims to complete API calls from a specific third-party library by extracting API call sequence~\cite{zhong2009mapo}. Variable name completion recommends variable references from declared variables, using data-flow graphs~\cite{allamanCode2Graph} or pointer network~\cite{li2018codewithPointer}. Arbitrary token completion aims to complete arbitrary tokens in the code sequence or the abstract syntax tree (AST) of code, using Tree-based NN~\cite{treeLstm,wang2022tree}, Graph NN~\cite{allamanCode2Graph,liuretrieval} or Transformer~\cite{codex}. For example, CodeFill~\cite{izadi2022codefill} improves GPT-2~\cite{svyatkovskiy2020intellicode} by predicting the code type when completing the next token, while Grammformer~\cite{guo2021learning} generates code completions with ``holes'' inserted in places where the model is uncertain. Our approach belongs to completing arbitrary token in code sequence. 

Nowadays, many works begin to focus on code completion in specific domains, such as test code~\cite{dinella2022togaTestCompletion,nie2023DeepTestCompletion}  or repository level code completion~\cite{project_miner,zhang2023repocoder}. Our approach focuses on domain adaptive code completion, which is a generic approach that can adapt to different domains by switching the database. It does not require access to the parameters of the LM, hence can be used with black-box LLMs like GPT-4.
\subsection{Retrieval-augment Language Model}
Retrieval-augmented language models (R-LMs) utilize retrieval-based techniques to improve the performance of LMs and can be divided into two main categories: block R-LMs and token R-LMs. Block R-LMs~\cite{izacard2022few, ReACC, guu2020retrieval} are similar to one-shot or few-shot learning~\cite{wang2020generalizing}, where one or a few examples are retrieved from a database instead of being randomly selected. Token R-LMs~\cite{knnlm,knnmt,meng2022gnn} retrieve tokens from database and then combine the retrieved results into the LM. Compared with block R-LMs, token R-LMs can update retrieval results at the same time of generating new tokens, hence our approach uses the architecture of token R-LM. However, token R-LMs suffer from high storage costs and require hyper-parameters selection to combine the inference results from the database and language model.

Various approaches have been proposed to address the limitations of token R-LMs. For example, $k$NN-Adapter~\cite{huang2023k} uses a trained network to determine the combination weights. To reduce the search cost, RetoMaton~\cite{alon2022neuro} uses the automaton states to save search time, while AdaptRet~\cite{he2021efficient} uses a trained network to decide whether to use the retrieval module. GNN-LM~\cite{meng2022gnn} selects similar texts and builds a contextual graph to incorporate into the language model. In contrast, our proposed approach does not require training or the addition of an additional module. By decoupling the datastore and language model, we can save the storage cost and utilize Bayesian inference to select suitable hyper-parameters at the same time.

\section{Conclusion}
In this paper, we propose $k$NM-LM for domain adaptive code completion. By utilizing the in-domain code to construct the retrieval database, the language model can adapt to complete code in target domain without fine-tuning. Specifically, $k$NM-LM builds a decoupled domain database (saving only tokens that the langauge model cannot predict correctly) and employs Bayesian inference to combine the results from the language model and the database. Experiments have shown that $k$NM-LM achieves the best results on intra-project and intra-scenario code completion tasks. 
Notably, our approach does not require access to the weights in the language model, nor does it require adding any additional neural network modules, making it feasible for leveraging the black-box language models on other domain-intensive code-related tasks, such as code summarization~\cite{hu2018deep,liuretrieval}, bug identification~\cite{zhou2019devign}, localization~\cite{niu2023rat} and repair~\cite{zhong2022standup4npr}. To these ends, we make all our code, data, and models publicly available.
\section{ACKNOWLEDGMENTS}
We are grateful to the anonymous reviewers for their useful comments and suggestions. This work was supported by National Key Research and Development Program of China (2022YFF0711404), Natural Science Foundation of Jiangsu Province, China (BK20201250), Cooperation Fund of Huawei-NJU Creative Laboratory for the Next Programming and CCF-Huawei Populus Grove Fund. Jidong Ge and Shangqing Liu are the corresponding authors.

\bibliographystyle{IEEEtran}
\bibliography{main}



\end{document}